\documentclass[%
 reprint,
 amsmath,amssymb,
 aps,superscriptaddress
]{revtex4-2}

\usepackage{graphicx}% Include figure files
\usepackage{dcolumn}% Align table columns on decimal point
\usepackage{bm}% bold math

\usepackage{epsfig}
\usepackage[english]{babel}
\usepackage{latexsym}
\usepackage{graphics}
\usepackage{subfigure}
\usepackage{graphicx}
\usepackage{amsmath}
\usepackage{hyperref}
\usepackage{amssymb}
\usepackage{color}
\begin{document}

\preprint{APS/123-QED}

\title{Single-photon ionization of H$_2^+$ in near-circular laser fields with lower photon energy}
\author{Z. Y. Chen}
\homepage{These authors contribute equally to this paper.}
\affiliation{College of Physics and Information Technology and Quantum Materials and Devices Key Laboratory
of Shaanxi Province's High Education Institution, Shaan'xi Normal University, Xi'an, China}
\author{S. Q. Shen}
\homepage{These authors contribute equally to this paper.}
\affiliation{College of Physics and Information Technology and Quantum Materials and Devices Key Laboratory
of Shaanxi Province's High Education Institution, Shaan'xi Normal University, Xi'an, China}
\author{M. Q. Liu}
\affiliation{College of Physics and Information Technology and Quantum Materials and Devices Key Laboratory
of Shaanxi Province's High Education Institution, Shaan'xi Normal University, Xi'an, China}
\author{J. Y. Che}
\affiliation{College of Physics, Henan Normal University, Xinxiang, China}
\author{Y. J. Chen}
\email{chenyjhb@gmail.com}
\affiliation{College of Physics and Information Technology and Quantum Materials and Devices Key Laboratory
of Shaanxi Province's High Education Institution, Shaan'xi Normal University, Xi'an, China}

\date{\today}

\begin{abstract}

We study single-photon ionization of aligned H$_2^+$ in low-intensity near-circular laser fields with lower photon energy numerically and analytically. The photoelectron momentum distribution (PMD) within the laser polarization plane, obtained by numerical simulations, shows a remarkable offset angle, which changes with changing the internuclear distance and the laser frequency. This phenomenon is different from that observed in recent experiments [Science 370, 339 (2020)] which is related to the PMD along the propagation direction of the laser. This phenomenon holds even for H$_2^+$ with short-range Coulomb potentials but disappears for atoms, different from that observed in attoclock experiments. We show that the molecular Coulomb potential near the two atomic centers plays an important role here and theory models associated with more accurate continuum wave function  of the molecule are needed for reproducing this phenomenon. This phenomenon can be useful for ultrafast probing of molecules with high resolution of several attoseconds or even zeptoseconds.

\end{abstract}

\maketitle

\section{Introduction}

The advances in ultrashort and ultrastrong laser technology has provided opportunities to precisely measure and control electron motion within atoms and molecules. The interaction of strong near-infrared lasers with atoms and molecules results in many interesting ultrafast physics phenomena, such as above-threshold ionization (ATI) \cite{Agostini1979, Yang1993, Paulus1994, Lewenstein1995, Becker2002}, non-sequential double ionization (NSDI) \cite{Niikura2003, Zeidler2005, Becker2012}, and high-order harmonic generation (HHG) \cite{McPherson1987, Huillier1991, Corkum1993, Lewenstein1994}.
These phenomena are further utilized in ultrafast probing procedures, through which the electron motion at its natural time scale of attosecond can be accessed \cite{Corkum2007,Krausz2009,Schultze2010}.

One of these procedures is the attoclock \cite{Eckle2008} which uses the ATI generated in a strong elliptically-polarized laser field (EPL) with high ellipticity and near-infrared wavelength to probe the attosecond tunneling dynamics of electron within atoms and molecules. Specifically, the offset angle in the photoelectron momentum distribution (PMD) of ATI is measured in attoclock. Then through constructing a mapping between the offset angle (which appears for atoms and molecule with long-range Coulomb potential) and the tunneling time (i.e., the time spent by the tunneling electron when it goes through the barrier formed by the laser field and the Coulomb potential), one can deduce the tunneling time from this angle \cite{Hofmann2019}. According to the well known strong-field approximation (SFA) \cite{Lewenstein1995} with electron-trajectory theory \cite{Becker2002}, for a strong near-circular laser field, the rescattering effect related to the electron motion in the time scale of more than a laser cycle is weak, and therefore the PMD obtained in attoclock  mainly encodes the electron dynamics in one laser cycle (which is about several femtoseconds for a near-infrared laser pulse). As a result, the attoclock can achieve the attosecond resolution of electron dynamics with a femtosecond-scale driving laser pulse.
Experimental and theoretical studies in attoclock \cite{Pfeiffer2012,Boge2013,Landsman2013,Landsman,Torlina2015,Sainadh2019,Che2021,Che2023} have provided deep insights into the tunneling dynamics of the electron, revealing influences of different aspects on tunneling time, such as the tunnel exit position and tunnel exit velocity \cite{Yan2010,Li2014,Lai2015,Shvetsov2016,Xie2020}, the near-nucleus and far-nucleus Coulomb effects and the shape of the Coulomb potential \cite{Peng2024,Shen2024}, etc..

Considering the principle of attoclock mentioned above, it is natural to think that the measure of PMD, generated in a near-circular high-frequency laser field with a timescale of several tens of or a hundred attosecond for a laser cycle, can help one to probe electron dynamics with a high time resolution of zeptoseconds. Indeed, this has been performed in recent experiments \cite{Sven2020}, where the PMD of H$_2$ along the direction of the laser propagation is measured and a time delay of 247 zeptoseconds is resolved which has been attributed to the time spent by the light when it propagates from one nucleus to the other nucleus of the diatomic molecule. This zeptosecond-resolution ultrafast dynamics phenomenon has aroused widespread research interest in recent years \cite{Ivanov2021, Trabert2021, Klaiber2022}. When this zeptosecond-related phenomenon is probed from the PMD in a plane perpendicular to the laser-polarization plane, one may ask what can be probed from the PMD in the laser-polarization plane as performed in general attoclock experiments?

In this paper, we study single-photon ionization of aligned H$_2^+$ in a near-circularly polarized high-frequency laser field with lower photon energy through numerical solution of time-dependent Schr\"{o}dinger equation (TDSE) in both two-dimensional (2D) and three-dimensional (3D) cases. We focus on the PMD calculated along the laser-polarization plane. The calculated PMD shows a remarkable offset angle, which is not sensitive to the laser intensity but changes with changing the internuclear distance and the laser frequency. In particular, this angle disappears for a model atom with similar ionization potential to H$_2^+$ and holds for a model H$_2^+$ with short-range Coulomb-potential, in contrast to phenomena observed in attoclock experiments performed in a near-circular near-infrared laser pulse. In general attoclock, the PMDs of both atoms and molecules with long-range Coulomb potential show a nonzero offset angle which, however, disappears for atoms and molecules with short-range Coulomb potential. Therefore, the offset angle for aligned H$_2^+$ in single-photon ionization studied here may be related to a new mechanism. As this angle can also not be reproduced by theoretical models which do not fully consider near-nucleus Coulomb effects on continuum electrons, we anticipate that this angle is closely related to the structure of the Coulomb potential near these two atomic nucleus of the molecule.
This angle can serve as a characteristic observable for exploring time-resolved dynamics of single-photon ionization of aligned molecules.

\section{Theory}
\subsection{Numerical methods}
Numerically, we first solve the TDSE in 2D cases to explore the single-photon ionization of H$_2^+$ and model atom in low-intensity and high-frequency elliptical laser fields with high ellipticity. We consider the case of the molecular axis aligned along the main axis  $\textbf{e}_x$ of the polarization ellipse. In the length gauge, the Hamiltonian of the model H$_2^+$ system studied here has the form of (atomic units of $\hbar=e=m_e=1$ are used throughout the paper unless stated otherwise)
\begin{equation}
H(t)=H_0+{\textbf{r}}\cdot\textbf{E}(t).
\end{equation}
Here, $H_0=\textbf{p}^2/2+V(\textbf{r})$ is the field-free Hamiltonian, and $V(\textbf{r})$ represents the Coulomb potential. For 2D cases \cite{Zuo, chen2013, lein1, Nalda}, we use the soft-core Coulomb potential  which can be expressed as $V(\mathbf{r})=-Ze^{-\rho\mathbf{r}_1}/\sqrt{\xi+\mathbf{r}_1^{2}}-Ze^{-\rho\mathbf{r}_2}/\sqrt{\xi+\mathbf{r}_2^{2}}$
with $\mathbf{r}_{1(2)}^2=(x\pm \frac{R}{2})^2+y^2$. In the above expression, the term $\rho$ is the screening parameter with $\rho=0$ for the long-range potential and $\rho=0.5$ for the short-range one. The term $R$ is the internuclear distance and $\xi=0.5$ is the smoothing parameter. The term $Z$ is the effective nuclear charge which is adjusted in such a manner that the ionization potential $I_p$ of H$_2^+$ reproduced here is $I_p=1.1$ a.u. for different internuclear distances R. Here, for comparison, the ionization potential of the system has been kept at a constant value at different R. For example, the parameters $Z$ used are $Z=1$ for H$_2^+$ with $R=2$ a.u. and $Z=0.85$ for model atom with $R=0$ a.u..

For the EPL case, the electric field $E(t)$ in Eq. (1) takes the form of $\mathbf{E}(t)=f(t) [{\mathbf{e}}_{x}E_{x}(t)+{\mathbf{e}}_{y}E_{y}(t)]$, where $E_{x}(t)=E_{1}\sin(\omega t)$, $E_{y}(t)=E_{2}\cos(\omega t)$. Here, ${\mathbf{e}}_{x}$ and ${\mathbf{e}}_{y}$ are the unit vector along the $x$ and $y$ axes. The amplitudes $E_{1}$ and $E_2$
are given by $E_{1}={E_0}/{\sqrt{1+\epsilon^2}}$ and $E_2=\epsilon {E_0}/{\sqrt{1+\epsilon^2}}$.
Here, $E_0$ is the maximal laser amplitude associated with the peak intensity $I$, $\omega$ is the laser frequency and $\epsilon=0.87$ is the ellipticity. The term $f(t)$ is the envelope function. We
use trapezoidally shaped laser pulses with a total duration of 400 cycles, which are linearly turned on and off for 50 optical cycles, and then kept at a constant intensity for 300 additional cycles. We solve the TDSE of $i\dot{\psi}(\mathbf{r},t)=H(t)\psi(\mathbf{r},t)$ in 2D cases using the spectral method \cite{Feit1982} with a time step of $\Delta t=0.05$ a.u.. We use a grid size of $L_x \times L_y=819.2 \times 819.2$ a.u. with space steps of $\Delta x=\Delta y=0.4$ a.u.. To avoid the reflection of the electron wave packet from the boundary and obtain the momentum-space wave function, the coordinate space is divided into inner and outer regions. This division is achieved by using a mask function $F(\mathbf{r})$ to separate the wave function ${\Psi}(\textbf{r},t)$ into the inner component ${\Psi}_{\rm in}(\textbf{r},t)$ and the outer component ${\Psi}_{\rm out}(\textbf{r},t)$ with ${\Psi}(\textbf{r},t)={\Psi}_{\rm in}(\textbf{r},t)+{\Psi}_{\rm out}(\textbf{r},t)$. The mask function
has the form of $F(\mathbf{r})=F_1(x)F_2(y)$. Here, $F_1(x)=\cos^{1/4}[\pi(|x|-r_x)/(L_x-2r_x)]$ for $|x|\geq r_x$ and $F_1(x)=1$ for $|x|< r_x$ with $r_x = 310$ a.u. being the absorbing boundary. The form of $F_2(y)$ is similar to $F_1(x)$. More details for solving the TDSE and obtaining the PMD can be found in Ref. \cite{Wang2017, Che2023}. Then we find the local maxima of the PMD and the offset angle $\theta$ is obtained with a Gaussian fit of the angle distribution of local maxima. Relevant 2D-TDSE results are presented in Fig. 1 and Fig. 5.

Simultaneously, we extended our simulations to 3D cases for H$_2^+$ with different $R$. Similar to 2D simulations, we also assume that the molecular axis is along the $x$ axis direction. The form of the coulomb potential $V(\mathbf{r})$ for 3D cases is similar to that for 2D cases, but the $z$ coordinate is taken into account. 
The grid size used here is $L_x\times L_y\times L_z=358.4\times 358.4\times 51.2$ a.u., with $\Delta x=\Delta y=0.7$ a.u. and $\Delta z=0.8$ a.u.. The time step used is $\Delta t=0.05$ a.u.. The mask function used here is $F(\mathbf{r})=F_1 (x)F_2(y)F_3(z)$. The term $F_{1}(x)$ ($F_{2}(y)$) used in 3D is similar to that used in 2D, but $r_x=140$ a.u. ($r_y=140$ a.u.). The expression of $F_{3}(z)$  is also similar to $F_{1}(x)$ but $r_z=19.2$ a.u. with $r_z$ being the absorbing boundary along the z direction. The PMD with respect to  $(p_x,p_y )$ is obtained with the integral of the 3D momentum distribution for the component of $p_z$. The corresponding 3D-TDSE results are shown in Fig. 6.

\subsection{Analytical description}
In this section, we discuss theory models used in analytical studies of single-photon ionization.

\textit{SFA}. Firstly, we introduce the SFA \cite{Keldysh1965,Faisal1973,Reiss1980,Lewenstein1995} which neglects the Coulomb effect on the continuum electron. Although it is generally considered that the SFA is applicable for tunneling ionization of atoms and molecules induced by a high-intensity and low-frequency laser field when  the Keldysh parameter $\gamma=\omega\sqrt{2I_p}/E_0$ is smaller than the unity \cite{Keldysh1965}, the SFA also holds for single-photon ionization of atoms and molecules induced by a low-intensity and high-frequency laser field \cite{Che2022}.
According to the SFA, the ionization amplitude $c(\textbf{p})$ for the photoelectron with the drift momentum $\textbf{p}$ in an EPL field can be written as
\begin{equation}
\begin{split}
c(\textbf{p})= & -i\int^{T_{p}}_0dt^\prime E_{x}(t'){\mathbf{e}}_{x}\cdot{\textbf{d}_i}{(\textbf{p}+\textbf{A}(t^\prime))}e^{iS(\textbf{p},t^\prime)} \\ &
-i\int^{T_{p}}_0dt^\prime E_{y}(t'){\mathbf{e}}_{y}\cdot{\textbf{d}_i}{(\textbf{p}+\textbf{A}(t^\prime))}e^{iS(\textbf{p},t^\prime)}.
\end{split}
\end{equation}
Here, $S(\textbf{p},t')=\int_{}^{t'}\{{[\textbf{p}+\textbf{A}(t''})]^2/2+I_p\}dt''$ is the semiclassical action and $T_{p}$ is the length of the total pulse.
The term $\textbf{d}_i(\textbf{p}+\textbf{A}(t'))=\langle{\textbf{p}+\textbf{A}(t'})|\textbf{r}\vert{{0}\rangle}$ denotes the dipole matrix element for the bound-free transition. The term $\textbf{A}(t)=-\int^{t}\textbf{E}(t')dt'$ is the vector potential of the electric field $\textbf{E}(t)$. Considering the case of single-photon ionization with $\omega>I_p$, one can write the electric field $\textbf{E}(t)$  as $E_{x}(t)=E_1\sin\omega t=E_1(e^{i\omega t}-e^{-i\omega t})/2i$ and $E_{y}(t)=E_2\cos\omega t=E_2(e^{i\omega t}+e^{-i\omega t})/2$. Inserting the above expressions of $\textbf{E}(t)$ into Eq. (2), the following approximate expression can be obtained
\begin{equation}\label{eq2}
\begin{split}
c(\textbf{p}) &\propto\dfrac{1}{2}{\int^{T_{p}}_0dt^\prime{E_{1}{\mathbf{e}}_{x}\cdot{\textbf{d}_i}{[\textbf{p}+\textbf{A}(t^\prime)]}e^{iS'(\textbf{p},t')}}} \\ &-\dfrac{i}{2}\int^{T_{p}}_0dt^\prime{E_{2}{\mathbf{e}}_{y}\cdot{\textbf{d}_i}{[\textbf{p}+\textbf{A}(t^\prime)]}e^{iS'(\textbf{p},t')}}, 
\end{split}
\end{equation}
with $S'(\textbf{p},t')= S(\textbf{p},t')-\omega t'$. In Eq. (3), the terms related to $e^{i(S(\textbf{p},t')+\omega t^\prime)}$ which corresponds to the fast oscillation in the integral have been neglected from the perspective of rotating-wave approximation.
The temporal integral in Eq. (3) can be also evaluated by the saddle-point method \cite{Lewenstein1995}, with solving the following equation
\begin{equation}
[\textbf{p}+\textbf{A}(t_s)]^2/2=\omega-I_p.
\end{equation}
Because of $\omega>I_p$, the solution $t_s\equiv t_0$ for Eq. (4) is real. It denotes the moment at which the electron is born through single-photon ionization. The solution shows the periodicity. That is to say, $t_0$ and $t_0+T$ correspond to the same momentum $\textbf{p}$. Here, $T=2\pi/\omega$ is the laser cycle. 
 We also call the momentum-time pair $(\textbf{p},t_{0})$ the electron trajectory.  The corresponding amplitude for the trajectory $(\textbf{p},t_{0})$ can be written as
\begin{equation}
\begin{split}
F(\textbf{p},t_0) &\propto\beta\lbrace[\dfrac{1}{2}E_1{\mathbf{e}}_{x} \cdot \textbf{d}_i(\textbf{p}+\textbf{A}(t_0))\\&-\dfrac{i}{2}E_2 {\mathbf{e}}_{y}\cdot \textbf{d}_i(\textbf{p}+\textbf{A}(t_0))]e^{iS'(\textbf{p},t_0)}\rbrace
\end{split}
\end{equation}
with $\beta\equiv({1/detA2})^{1/2}$. The term $detA2$ in the definition of $\beta$ is the determinant of the matrix formed by the second derivatives of $S'$ \cite{Lewenstein1995}.
The whole amplitude for photoelectron with a momentum $\textbf{p}$ can be written as
\begin{equation}
c(\textbf{p})\propto \sum_{t_s} F(\textbf{p},t_s).
\end{equation}
The sum runs over all possible saddle points $t_s$.

\textit{Expressions of dipoles with PWA}. With linear combination of atomic orbitals-molecular orbitals (LCAO-MO) approximation for describing the bound state $|0\rangle$ and the plane wave approximation (PWA) for describing the continuum state $|\textbf{v}\rangle$ \cite{chen2009}, the dipole of H$_2^+$ can be written as
\begin{equation}\label{eq1}
\begin{split}
{\textbf{e}}_x\cdot \textbf{d}_i(\textbf{v})=\cos (\textbf{v}\cdot \frac{\textbf{R}}{2})({\textbf{v}}\cdot \vec{\textbf{e}}_x)\frac{32\pi\kappa}{(\kappa^2+\textbf{v}^2)^3},
\end{split}
\end{equation}
\begin{equation}\label{eq1}
\begin{split}
{\textbf{e}}_y\cdot \textbf{d}_i(\textbf{v})=\cos (\textbf{v}\cdot \frac{\textbf{R}}{2})({\textbf{v}}\cdot \vec{\textbf{e}}_y)\frac{32\pi\kappa}{(\kappa^2+\textbf{v}^2)^3},
\end{split}
\end{equation}
with $\kappa=\sqrt{2I_p}$.

\textit{CWA for dipoles}. As discussed in \cite{chen2009}, due to the Coulomb effect, the continuum-state wave function of $H_0$ near the position of the nucleus differs remarkably from the description of PWA $\langle{\textbf{r}} | \textbf{v} \rangle \propto e^{i\textbf{v} \cdot \textbf{r}}$ with energy $E_v=\textbf{v}^2/2$. A Coulomb-corrected PWA (which is called CWA here for simplicity) can be written as  $\langle{\textbf{r}} | \textbf{v} \rangle \propto e^{i\textbf{v}_k \cdot \textbf{r}}$ with the assumptions of  $H_0 | \textbf{v} \rangle=\textbf{v}^{2}/2 | \textbf{v} \rangle$, $\textbf{v}^2_k/2=\textbf{v}^2/2+I_p$  and $\textbf{v}_k/{v}_k=\textbf{v}/{v}$. The above expression uses the effective momentum $\textbf{v}_k$ which considers the influence of the ionization potential related to the Coulomb potential on the momentum $\textbf{v}$ of the plane wave. Using this expression, Eq. (7) and Eq. (8) can be written as
\begin{equation}\label{eq1}
\begin{split}
{\textbf{e}}_x\cdot \textbf{d}_i(\textbf{v})=\cos (\frac{v_{k}R}{2}\cos\theta')\cos{\theta_{0}}\frac{32\pi\kappa v_k}{(\kappa^2+v^2)^3},
\end{split}
\end{equation}
\begin{equation}\label{eq1}
\begin{split}
{\textbf{e}}_y\cdot \textbf{d}_i(\textbf{v})=\cos (\frac{v_{k}R}{2}\cos\theta')\sin{\theta_{0}}\frac{32\pi\kappa v_k}{(\kappa^2+v^2)^3}.
\end{split}
\end{equation}
Here, $v =|\textbf{v}|$ and $v_k =|\textbf{v}_k|$. The term $\theta'$ denotes the angle between the momentum $\textbf{v}$ and the molecular axis, and the term $\theta_{0}$ denotes the angle between the momentum $\textbf{v}$ and the vector ${\mathbf{e}}_{x}$. In Ref. \cite{Che2022}, it has been shown that Eqs. (9) and (10) give an applicable description of the effect of two-center interference in single-photon ionization of H$_2^+$ induced by a linearly polarized high-frequency laser field with lower photon energy.

\textit{SCC}. One can also use the single-center continuum (SCC) wave function to describe or approximately describe the continuum electrons of an atom or a molecule \cite{Joulakian1996,Ciappina2007}, which can be written as
\begin{equation}
\begin{split}
\langle{\textbf{r}} |\textbf{v} \rangle=\varphi^{SCC}(\textbf{v},\textbf{r}) = \phi(\textbf{v},\textbf{r}) C(\textbf{v},\textbf{r}),
\end{split}
\end{equation}
where $C(\textbf{v},\textbf{r})=N(\nu)_{1}F_{1}[-i\nu,1,-ivr-i\textbf{v}\cdot\textbf{r}]$, $\phi(\textbf{v},\textbf{r})=e^{i\textbf{v}\cdot\textbf{r}}/{(2\pi)^{3/2}}$, and $\textbf{r}$ is the position of the electron with respect to the center of mass of the system. Here, $N(\nu)=\exp(\pi\nu/2)\Gamma(1+i\nu)$ is the usual Coulomb normalization factor, $\nu=1/v$ expresses the Sommerfeld parameter, and $_{1}F_{1}$ is the confluent hypergeometric function.

\textit{TCC}. Besides the SCC, a two-center continuum (TCC)  wave function \cite{Joulakian1996,Ciappina2007}, which considers the two-center characteristic of the molecular Coulomb potential, can also be used to approximately describe the continuum state of the molecule. The TCC can be expressed as
\begin{equation}
\begin{split}
\langle{\textbf{r}} |\textbf{v} \rangle=\varphi^{TCC}(\textbf{v},\textbf{r},\textbf{R}) = \phi(\textbf{v},\textbf{r}) C(\textbf{v},\textbf{r}_{1})C(\textbf{v},\textbf{r}_{2}),
\end{split}
\end{equation}
where $\textbf{r}_{1,2}=\textbf{r}\pm\textbf{R}/2$ are the positions of the electron with respect to the two nuclei.
By inserting the expressions of SCC and TCC into Eq. (5) to calculate the bound-free dipole transition matrix element, we can also obtain the Coulomb-corrected ionization amplitude $c(\textbf{p})$ through Eq. (6) for atoms and molecules.

It is worth noting that the CWA considers only the correction of the Coulomb potential on the momentum of the continuum wave function near the atomic nuclei of the molecule. By comparison, the SCC (TCC) is  the exact limit of the outgoing wave in the field of the nucleus of the atom (the two nuclei of the molecule) in the asymptotic limit of $r\rightarrow\infty$ ($r_1,r_2\rightarrow\infty$) \cite{Joulakian1996,Ciappina2007}. The application of PWA, CWA, SCC and TCC for describing single-photon ionization of H$_2^+$ and model atom in near-circular EPL field with lower photon energy is explored in Fig. 2.

\section{Results and discussions}
%%%%%%%%%%%%%%%%%%%%%%%%%%%%%%%%%%%%%%%%%%%%%%%%%%%%%%%%%%%%%%%%%%%%%%%%%%%%%%%%%%%%%%%%
\begin{figure}[t]
\begin{center}
\rotatebox{0}{\resizebox *{8.6cm}{7.2cm} {\includegraphics {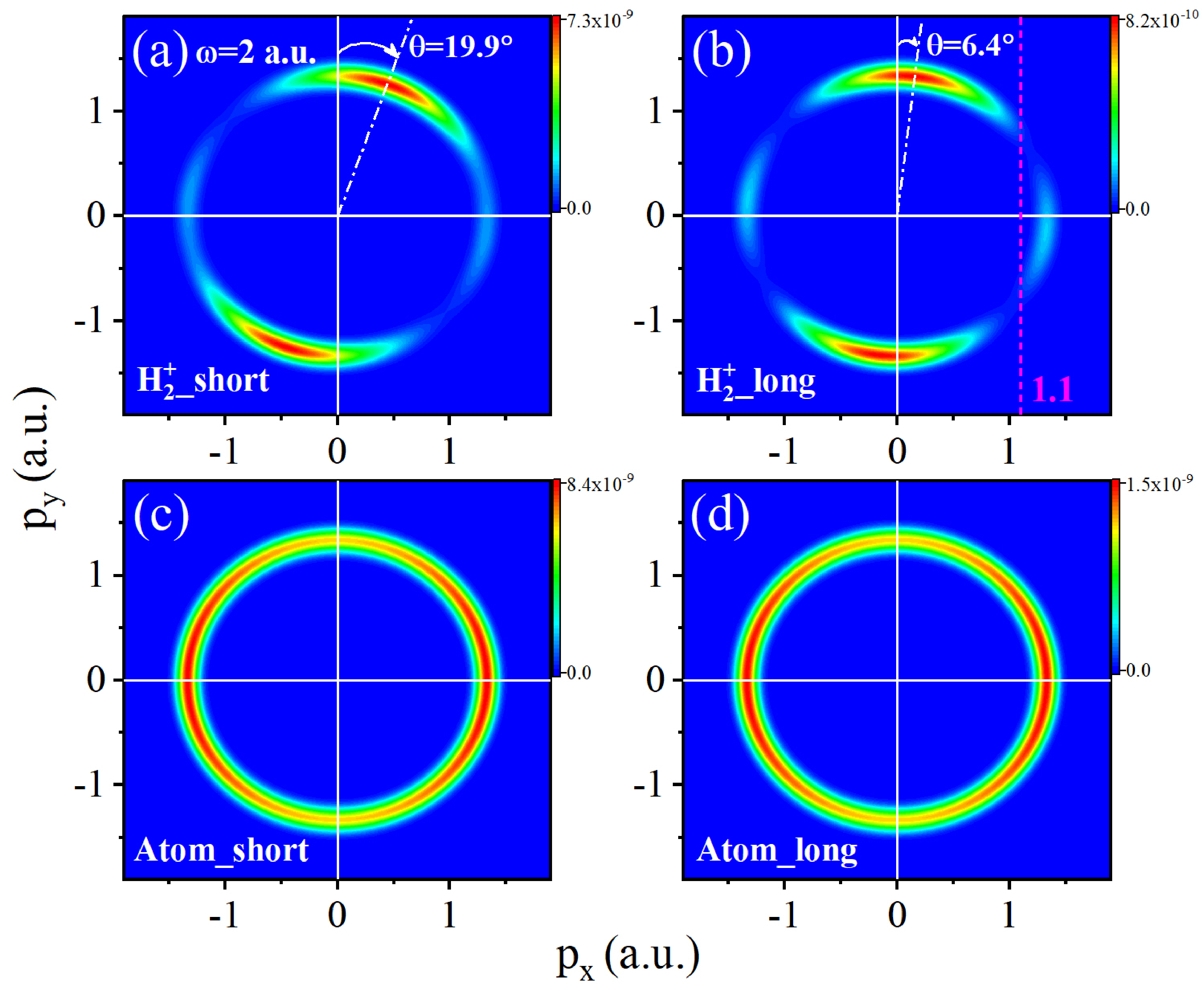}}}
\end{center}
\caption{PMDs of H$_{2}^{+}$ with $ R=2 $ a.u. and model atom obtained with 2D-TDSE in low-intensity high-frequency EPL laser fields. (a) and (c): H$_{2}^{+}$ and model atom with the short-range potential; (b) and (d): H$_{2}^{+}$ and model atom with the long-range potential. The laser parameters used here are $I=1\times10^{13}$W/cm$^{2}$ and $\omega=2$ a.u.. The white dash-dotted lines in (a) and (b) indicate the offset angles of the PMD. The red dashed line in (b) indicates the position of the interference minimum, and the red number denotes the corresponding momentum $p_x$.}
\label{fig1}
\end{figure}
%%%%%%%%%%%%%%%%%%%%%%%%%%%%%%%%%%%%%%%%%%%%%%%%%%%%%%%%%%%%%%%%%%%%%%%%%%%%%%%%%%%%%%%%%%
\textit{Cases of 2D TDSE}. In Fig. 1, we first present the results for H$_{2}^{+}$ with $R=2$ a.u. and model atom with short-range and long-range Coulomb potentials obtained by 2D-TDSE in high-frequency EPL fields. Here, the laser parameters used are $I=1\times10^{13}$W/cm$^{2}$ and $\omega=2$ a.u.. For the convenience of comparing with model results for which one generally first presents the Coulomb-free SFA results and then shows the Coulomb-corrected ones, we present the TDSE results of short-range (long-range) potential in the left (right) column of Fig. 1.

%%%%%%%%%%%%%%%%%%%%%%%%%%%%%%%%%%%%%%%%%%%%%%%%%%%%%%%%%%%%%%%%%%%%%%%%%%%%%%%%%%%%%%%%
\begin{figure}[t]
\begin{center}
\rotatebox{0}{\resizebox *{7.8cm}{9.5cm} {\includegraphics {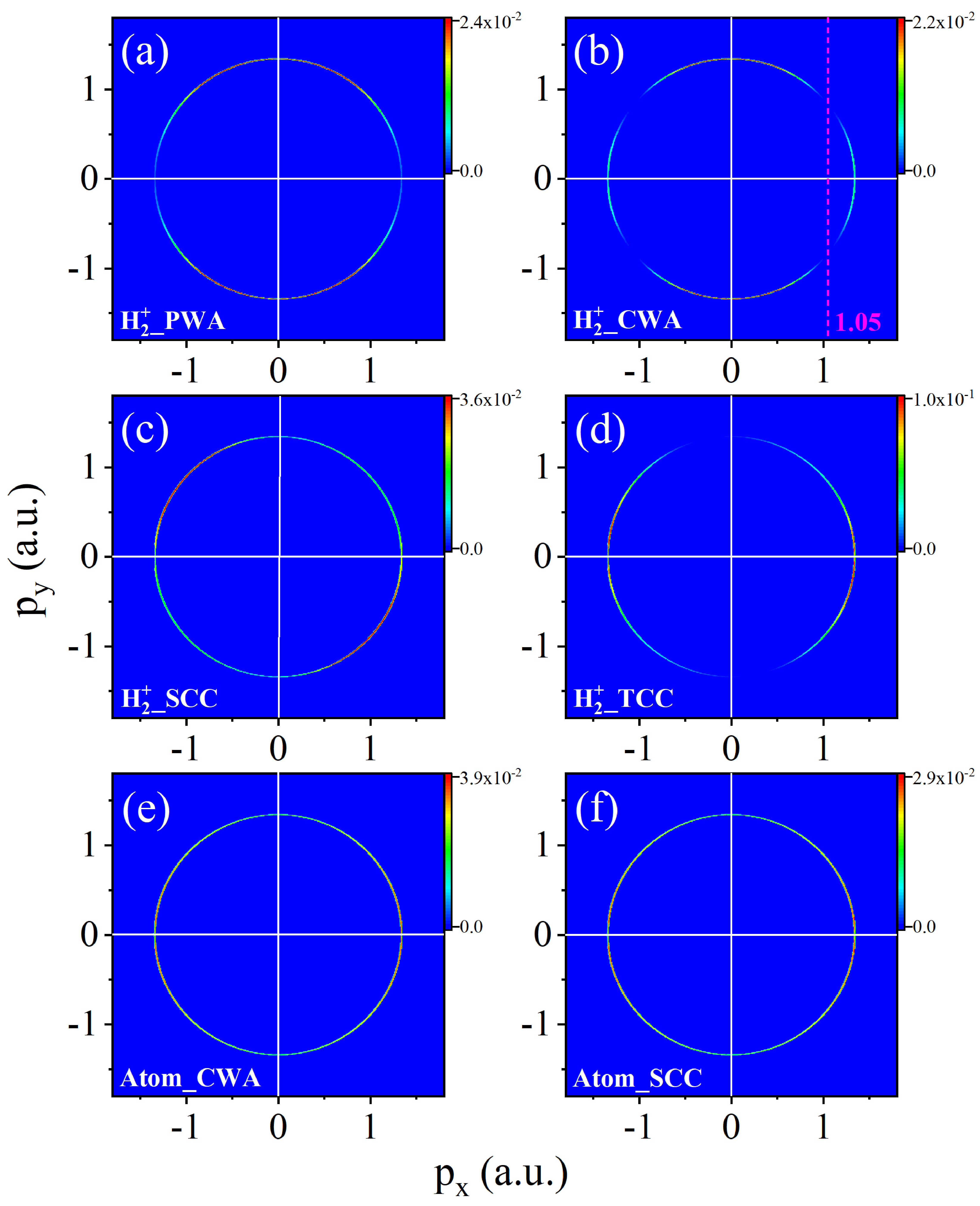}}}
\end{center}
\caption{PMDs of H$_{2}^{+}$ with $R=2$ a.u. (a-d) and model atom (e-f) obtained with different models in low-intensity high-frequency EPL laser fields. (a) : PWA; (b) and (e): CWA; (c) and (f): SCC and (d): TCC. The laser parameters used here and the red dashed line in (b) are as in Fig. 1.}
\label{fig2}
\end{figure}
%%%%%%%%%%%%%%%%%%%%%%%%%%%%%%%%%%%%%%%%%%%%%%%%%%%%%%%%%%%%%%%%%%%%%%%%%%%%%%%%%%%%%%%%%%

\textit{Characteristic structures in PMD}. For the case of H$_2^+$ with long-range Coulomb potential, one can observe from Fig. 1(b) that the distribution shows two remarkable characteristics. Firstly, the distribution shows clear interference minima located at the positions of $|p_x|=1.1$ a.u., as indicated by the red dashed line. Secondly, the distribution shows a clockwise rotation with an offset angle of $\theta=6.4^o$ (the angle between the line connecting the origin and the momentum corresponding to the maximal amplitude in PMD and the axis of $p_x=0$ a.u.), similar to the case of attoclock, as indicated by the white dash-dotted line. The interference and rotation phenomena hold for H$_2^+$ with short-range Coulomb potential, as shown in Fig. 1(a). This offset angle in Fig. 1(a) is about $\theta=19.9^o$, larger than that in Fig. 1(b). The rotation phenomenon appearing in Fig. 1(a) of short-range potential is different from that in general attoclock where the rotation phenomenon appears only for cases of long-range potential. By comparison, the atomic cases of both short and long range potentials shown in Figs. 1(c) and 1(d) do not show the interference phenomenon and the rotation phenomenon. The interference phenomenon has been discussed in \cite{Che2022} in detail for single-photon ionization of H$_2^+$ in a linear laser field with lower photon energy. Here, we mainly focus on the rotation phenomenon, while using the interference phenomenon as an auxiliary means to identify the applicability of theoretical models.  
It should be mentioned that in \cite{Fernandez2009} which is focused on the photoelectron angular distribution for single-photon ionization of molecules in a circular laser field with higher photon energy, it has been shown that this interference effect is not easy to resolve and the rotation of the angular distribution is not remarkable for high-energy cases.

%%%%%%%%%%%%%%%%%%%%%%%%%%%%%%%%%%%%%%%%%%%%%%%%%%%%%%%%%%%%%%%%%%%%%%%%%%%%%%%%%%%%%%%%%
\begin{figure}[t]
\begin{center}
\rotatebox{0}{\resizebox *{8.6cm}{7.2cm} {\includegraphics {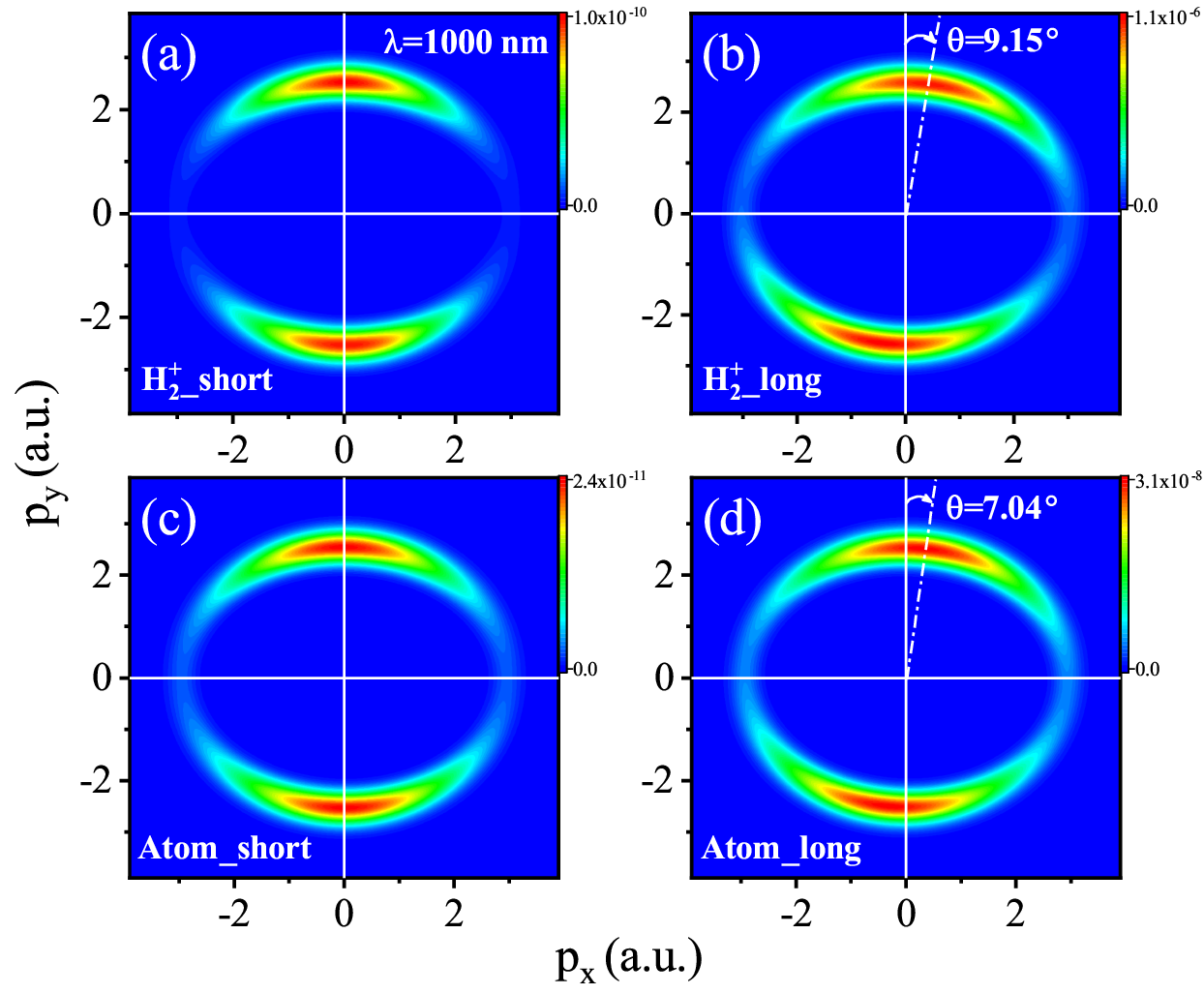}}}
\end{center}
\caption{Same as  Fig. 1 but corresponding to high-intensity low-frequency EPL laser fields. The laser parameters used here are $I=1\times10^{15}$W/cm$^{2}$ and $\lambda=1000$nm.}
\label{fig3}
\end{figure}
%%%%%%%%%%%%%%%%%%%%%%%%%%%%%%%%%%%%%%%%%%%%%%%%%%%%%%%%%%%%%%%%%%%%%%%%%%%%%%%%%%%%%%%%%

\textit{Model predictions}. To understand the TDSE results in Fig. 1, we show model predictions of Eq. (6) with different descriptions for the continuum wave function, as introduced in Sec. II. B. For H$_2^+$ shown in Figs. 2(a) to 2(d), one can observe that all of the PWA, CWA, SCC and TCC results differ remarkably from the corresponding TDSE results in Figs. 1(a) and 1(b). By comparison, the CWA result in Fig. 2(b) is somewhat nearer to the TDSE one. Specifically, the  CWA result in Fig. 2(b) shows clear interference minima which arise from two-center interference and are around the momenta of $|p_x|=1.05$ a.u. near to the TDSE result of $|p_x|=1.1$ a.u., as indicated by the red dashed line. However, the CWA result does not show the rotation of the PMD as the TDSE result does and has maxima around the axis of $p_x=0$ a.u.. The PWA result in Fig. 2(a) also shows maxima around $p_x=0$ a.u. but does not show the interference minima around $|p_x|=1.1$ a.u.. The SCC result in Fig. 2(c) is somewhat similar to the PWA result in Fig. 2(a) but shows a rotation of about $-45^o$ relative to the PWA result. The rotation relation between PWA and CWA PMDs is similar to that between PWA and CWA dipoles \cite{chen2013,Haessler2012}.
The TCC result in Fig. 2(d) shows maxima near the axis of $|p_y|=0$ a.u. and the interference minima around $|p_x|=1.1$ a.u. also do not appear in the TCC result. For the model atom with similar $I_p$ to H$_2^+$, all of predictions of PWA, CWA and SCC  are similar. In Figs. 2(e) and 2(f), we show the results of CWA and SCC which are also similar to the TDSE results in Figs. 1(c) and 1(d), showing maxima around the axis of $|p_y|=0$ a.u.. As the simulations with PWA, CWA, SCC and TCC for H$_2^+$ can not produce the interference and rotation phenomena as in TDSE, we anticipate that more precise Coulomb-included theory models which put an emphasis on the effect of the Coulomb potential around the two nuclei of the molecule on the continuum electron need to be developed to quantitatively explain the phenomena observed in Figs. 1(a) and 1(b).

%%%%%%%%%%%%%%%%%%%%%%%%%%%%%%%%%%%%%%%%%%%%%%%%%%%%%%%%%%%%%%%%%%%%%%%%%%%%%%%%%%%%%%%%%
\begin{figure}[t]
\begin{center}
\rotatebox{0}{\resizebox *{8.6cm}{7.2cm} {\includegraphics {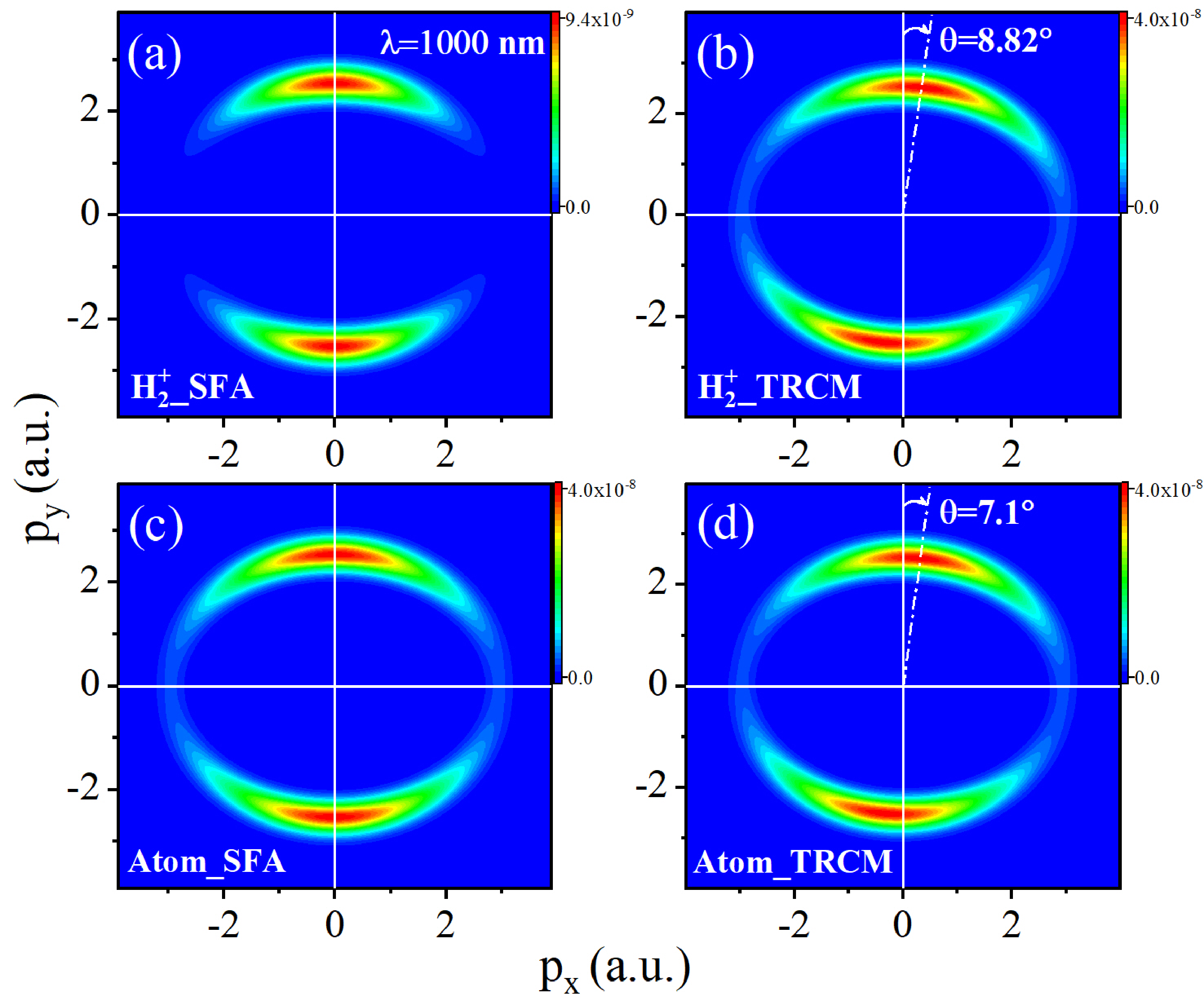}}}
\end{center}
\caption{Same as  Fig. 3 but obtained with SFA (the left column) and TRCM (right).}
\label{fig4}
\end{figure}
%%%%%%%%%%%%%%%%%%%%%%%%%%%%%%%%%%%%%%%%%%%%%%%%%%%%%%%%%%%%%%%%%%%%%%%%%%%%%%%%%%%%%%%%%

\textit{Comparisons to attoclock}. As a comparison, in Fig. 3, we show the attoclock results for H$_2^+$ and model atoms with both long-range and short-range potentials. As both the results of long-range potential for  the molecule and the atom show a rotation in Figs. 3(b) and 3(d), the results of short-range potential in Figs. 3(a) and 3(c) do not show the rotation, suggesting that the rotation phenomenon observed in attoclock for both atoms and molecules  is associated with the long-range potential. In Fig. 4, we further show the results of attoclock for atoms and molecules calculated using the SFA which does not consider  the Coulomb effect and a Coulomb-included model termed as tunneling-response-classic motion (TRCM) model which considers the effect of the Coulomb potential in the range from the nucleus to the tunnel exit \cite{Che2023,Peng2024,Shen2024}. The model results in Fig. 4 agree with the TDSE ones in Fig. 3, suggesting that the rotation phenomenon in attoclock originates from the near-nucleus Coulomb effect. With the above discussions on the origin of the rotation in attoclock, this rotation phenomenon observed in Fig. 1 for single-photon ionization of H$_2^+$ with both long-range and short-range potentials can be further understood as arising from the Coulomb effect around these two atomic centers of the molecule.

%%%%%%%%%%%%%%%%%%%%%%%%%%%%%%%%%%%%%%%%%%%%%%%%%%%%%%%%%%%%%%%%%%%%%%%%%%%%%%%%%%%%%%%%%
\begin{figure}[t]
\begin{center}
\rotatebox{0}{\resizebox *{5.5cm}{9cm} {\includegraphics {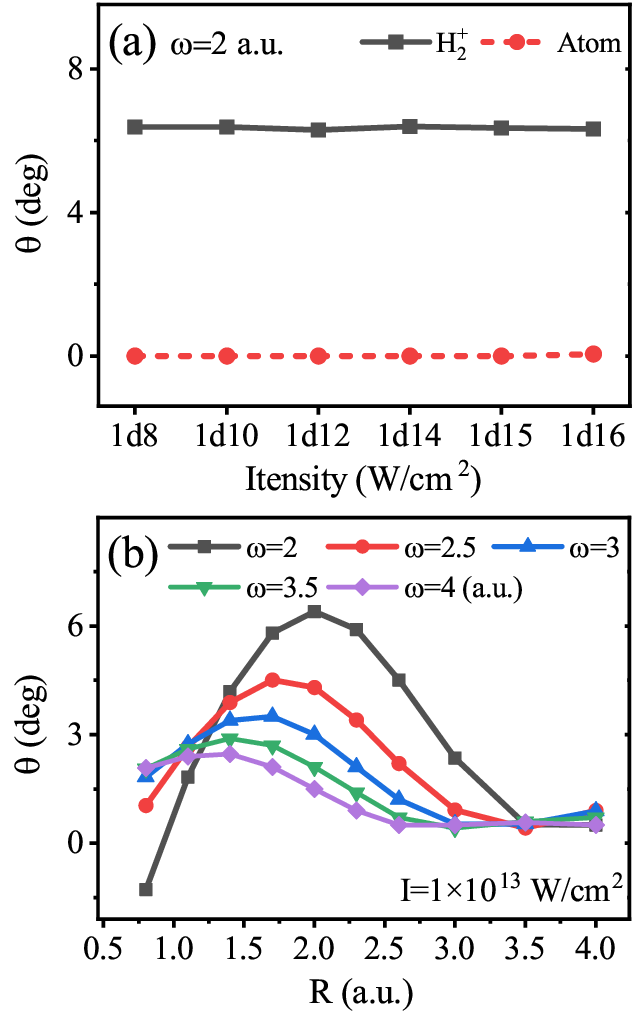}}}
\end{center}
\caption{Comparisons of the offset angle in PMD obtained with 2D-TDSE for different laser intensities $I$, internuclear distances $R$ and frequencies $\omega$ in high-frequency EPL fields. (a): Offset angles of H$_{2}^{+}$ with $R=2$ a.u. and model atom in the laser intensity range of $I=1\times10^{8}$W/cm$^{2}$ to $I=1\times10^{16}$W/cm$^{2}$ with $\omega=2$ a.u.. (b): Offset angles of H$_{2}^{+}$ at different $R$ and $\omega$.}
\label{fig5}
\end{figure}
%%%%%%%%%%%%%%%%%%%%%%%%%%%%%%%%%%%%%%%%%%%%%%%%%%%%%%%%%%%%%%%%%%%%%%%%%%%%%%%%%%%%%%%%%
\textit{Roles of laser and molecular parameters}. Next, we explore the dependence of the offset angle observed in Fig. 1(b) on laser and molecular parameters. Relevant results are presented in Fig. 5, where we show the TDSE offset angles related to long-range Coulomb potential as functions of the laser intensity (Fig. 5(a)), the laser frequency and the internuclear distance (Fig. 5(b)). Firstly, one can observe from Fig. 5(a) that this angle is not sensitive to the laser intensity $I$. Secondly, the results in Fig. 5(b) show that at a fixed laser frequency $\omega$, this angle first increases with the increase of the distance $R$, then this angle arrives at a maximum at a characteristic distance $R_c$. Thereafter, this angle becomes to decrease with increasing the distance $R$. This characteristic distance shifts towards larger distances $R$ for smaller frequencies $\omega$. In addition, when the distance is small such as  $R\leq1.1$ a.u., this angle seems to increases when increasing the frequency $\omega$. For a fixed distance located in the parameter region of $1.1<R<3.5$ a.u., this angle clearly increases with the decrease of the frequency. For the case of $R\geq3.5$ a.u., this angle has a small value and is not sensitive to $\omega$. Therefore, to observe the rotation phenomenon characterized by the offset angle in single-photon ionization, relatively smaller frequencies $\omega$ with the distance $R$ around the equilibrium separation of H$_2^+$ are preferred.

The potential mechanisms for the results in Fig. 5 may be as follows. Firstly, for smaller $R$, the molecule is nearer to an atom  which does not show this rotation phenomenon as seen in Fig. 1(d), and therefore for a fixed  frequency $\omega$, this angle is smaller for distances R smaller than the characteristic distance $R_c$. For distances $R$ larger than the characteristic distance $R_c$, the influence of one nucleus of the molecule on another nucleus decreases and therefore this angle also decreases. For higher laser frequencies (photon energy), the structure of the molecule plays a smaller role in single-photon ionization, and therefore this angle arising from the two-center characteristic of the molecular structure also decreases.

%%%%%%%%%%%%%%%%%%%%%%%%%%%%%%%%%%%%%%%%%%%%%%%%%%%%%%%%%%%%%%%%%%%%%%%%%%%%%%%%%%%%%%%%%
\begin{figure}[t]
\begin{center}
\rotatebox{0}{\resizebox *{8.6cm}{7.8cm} {\includegraphics {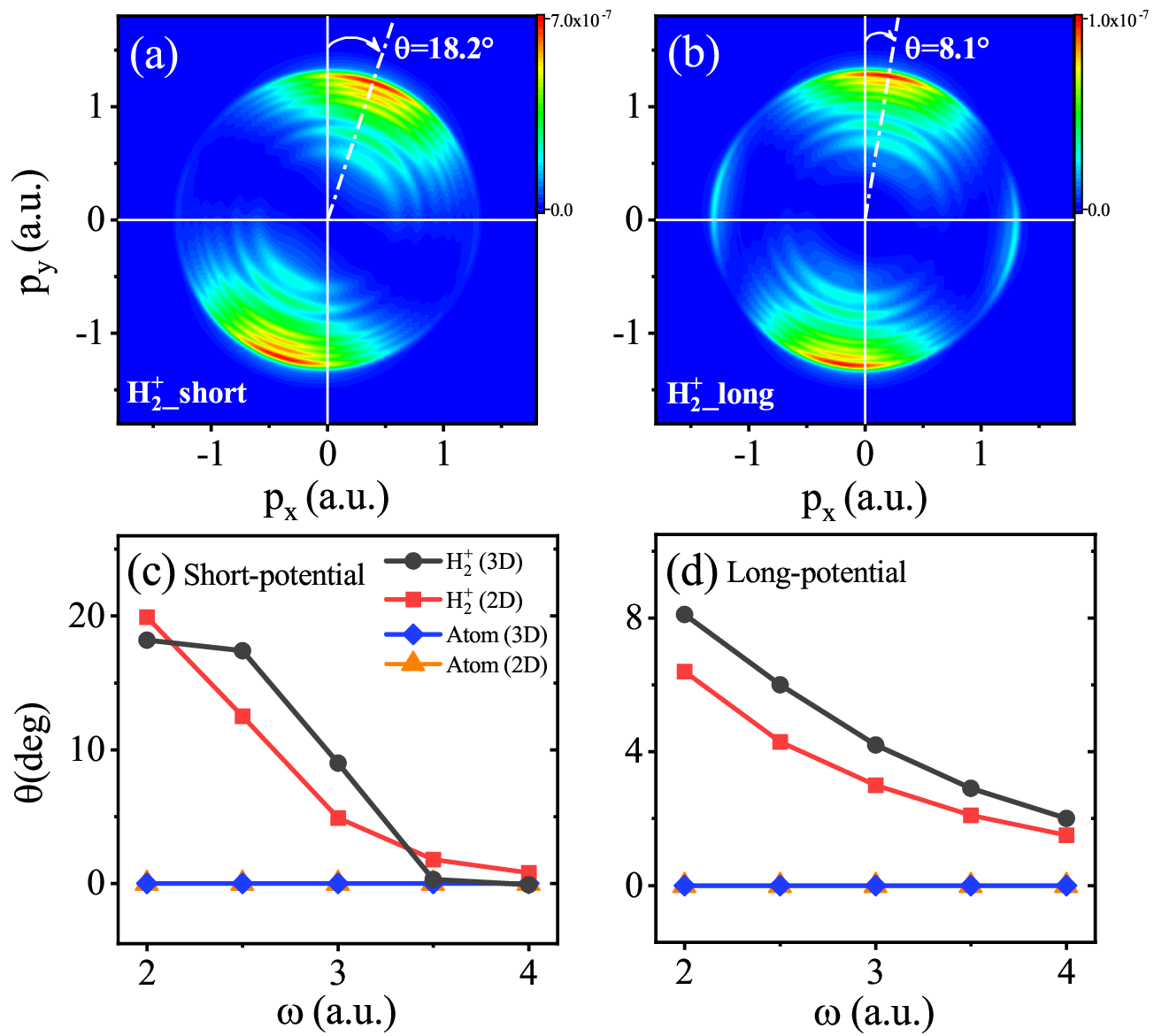}}}
\end{center}
\caption{PMDs of H$_{2}^{+}$ with $R=2$ a.u. obtained with 3D-TDSE for short-range (a) and long-range (b) Coulomb potentials in a low-intensity high-frequency EPL field with $I=1\times10^{13}$W/cm$^{2}$ and $\omega=2$ a.u.. In (c) and (d), we also compare the offset angles in PMDs for 2D and 3D cases of H$_2^+$ with $R=2$ a.u. and model atom with short-range (c)  and long-range (d) Coulomb potentials, obtained with different laser frequencies $\omega$ at $I=1\times10^{13}$W/cm$^{2}$.}
\label{fig6}
\end{figure}
%%%%%%%%%%%%%%%%%%%%%%%%%%%%%%%%%%%%%%%%%%%%%%%%%%%%%%%%%%%%%%%%%%%%%%%%%%%%%%%%%%%%%%%%%
\textit{Cases of 3D TDSE}. To check our results, we have also extended our simulations to 3D cases of H$_2^+$ and relevant results are shown in Fig. 6. The 3D distributions in the first row of Fig. 6 are similar to the 2D ones in the first row of Fig. 1. The PMDs of 3D H$_2^+$ with short-range and long-range Coulomb potentials in Figs. 6(a) and 6(b) both show the remarkable rotation phenomenon and the offset angles are also near to the corresponding 2D ones. In Figs. 6(c) and 6(d), we show the direct comparison of the offset angle between 2D and 3D cases of H$_2^+$ and model atom for short-range (Fig. 6(c)) and long-range (Fig. 6(d)) Coulomb potentials at different frequencies $\omega$. One can observe from Figs. 6(c) and 6(d) that for all of cases, the 3D results are similar to the 2D ones. Specifically, the 3D results show that the rotation phenomenon holds for H$_2^+$ with both long and short-range potentials and disappears for the atomic cases, suggesting that the rotation phenomenon is closely associated with the Coulomb potential around these two atomic centers. This offset angle related to the rotation phenomenon in 3D simulations decreases when increasing the frequency $\omega$ and is somewhat larger than the corresponding 2D prediction on the whole.

The rotation phenomenon may arise from the reason that for  H$_2^+$ with two-center Coulomb potential, due to the influence of one nucleus on another nucleus, the electron born through single-photon ionization can not respond to the change of the near-circular laser field immediately, whereas for model atom with single-center Coulomb potential, the electron does so. If we assume that the response time $\tau$ in single-photon ionization is also proportional to the offset angle $\theta$ with the relation $\theta\approx\omega\tau$ as in attoclock \cite{Che2023}, for the angle of $\theta=6.4^o$ in Fig. 1(b), we have $\tau\approx1.34$ attoseconds. In addition, the quantum interference related to the complex components of the actual two-center Coulomb continuum wave function may also play a role in the rotation phenomenon.
To verify the above assumptions, further detailed theory studies are needed on this issue. It is worth noting that quantum phenomena in attosecond science have also been a hot issue in recent research \cite{Cruz2024}.

\section{Conclusion}
In summary, we have studied single-photon ionization of H$_2^+$ in near-circular laser fields with lower photon energy which is not far larger than the ionization potential. The PMD calculated through TDSE shows a rotation phenomenon, somewhat similar to that observed in attoclock experiments related to tunneling ionization of atoms and molecules in strong near-circular near-infrared laser fields. The difference is that this phenomenon observed here holds for H$_2^+$ with both long-range and short-range Coulomb potentials and disappears for atoms. By comparison, the rotation phenomenon in attoclock appears for both atoms and molecules with long-range Coulomb potential but does not appear for short-range ones. The offset angle related to the rotation phenomenon here is not sensitive to the laser intensity, but depends strongly on the laser frequency and the internuclear distance $R$. On the whole, this angle is smaller for larger laser frequencies and larger distances $R$. Specifically, for a fixed laser frequency, this angle curve as a function of the internuclear distance is relatively smooth and shows a peak. The value of the peak is larger for smaller laser frequencies and the characteristic distance corresponding to the peak shifts towards smaller distances $R$ for larger frequencies $\omega$. As the rotation phenomenon observed here also appears for H$_2^+$ with short-range potential, one can conclude that this phenomenon is closely related to the effect of the Coulomb potential around these two atomic centers of the molecule. Therefore, this phenomenon can be useful for studying the characteristic of the molecular Coulomb potential near atomic nuclei. 
Due to the ultrashort property of the laser cycle  ($\sim$100 attoseconds for the present cases) where the ionization event occurs, 
this phenomenon  may also contribute to the study of time-resolved single-photon ionization dynamics of the molecule, with a high resolution of several attoseconds or even zeptoseconds.

\section{ACKNOWLEDGMENTS}
This work was supported by the National Natural Science Foundation of China (Grant No. 12174239), Shaanxi Fundamental Science Research Project for Mathematics and Physics (Grant No. 23JSY022).

\end{document}